\documentclass[
 reprint,
 amsmath,amssymb,
 aps,
]{revtex4-2}
\usepackage[ruled,vlined]{algorithm2e}
\usepackage{mathrsfs}
\usepackage{braket}
\usepackage{color}
\usepackage{float} 
\usepackage{graphicx}
\usepackage{dcolumn}
\usepackage{bm}
\usepackage{placeins}

\errorstopmode

\begin{document}

\preprint{APS/123-QED}

\title{Perfect Quantum State Revivals: Designing Arbitrary Potentials \\
with Specified Energy Levels}

\author{Aaron Danner}
 \email{adanner@nus.edu.sg}
\affiliation{Department of Electrical and Computer Engineering, National University of Singapore, 4 Engineering Drive 3, 117583, Singapore}

\author{Tom\'a\v{s} Tyc}
\affiliation{Department of Theoretical Physics and Astrophysics, Faculty of Science, Masaryk University, Kotl\'a\v{r}sk\'a 2, 61137 Brno, Czechia}

\date{\today}

\begin{abstract}
It is known that there exist a limited number of analytic potentials with the unusual property that any bound quantum state therein will be periodic in time.  This is known as a perfect quantum state revival. Examples of such potentials are the infinite well, quantum harmonic oscillator and the P\"{o}schl-Teller potentials; here, we present a general method of designing such potentials.  A key requirement is that their energy eigenvalues have integer spacings (up to a prefactor).  We first analyze the required conditions which permit quantum state revivals for potentials in general, and then we use techniques of iterated Hamiltonian intertwining to construct potentials exhibiting perfect quantum revivals. Our method can readily be extended to multiple dimensions. 
\end{abstract}

\maketitle

\section{\label{introduction}Introduction}

When a quantum wavepacket moving in a potential is launched, it spreads out, explores the potential, and in general never comes back to the exact initial state. However, under certain circumstances the wavepacket does recover to the initial state, and this almost magical effect is called \textit{quantum state revival}~\cite{ROBINETT_2004}. This happens in general if the energy of the wavepacket is restricted to a narrow interval~\cite{amjphys-kostelecky1996}, and the character of the energy spectrum determines the revival time. Wavepacket revivals have been investigated in different contexts, both theoretically and experimentally~\cite{PhysRevA.43.5153-rydberg_atoms,PhysRevA.109.063316}, including relativistic systems~\cite{prl104-120403-strange}, and other related effects such as fractional revivals and superrevivals can be observed too; a comprehensive review can be found in Ref.~\cite{ROBINETT_2004}.

At the same time, there exist certain quantum systems where the energy spectra meet particular requirements that allow exact wavepacket revival for {\em any} wavepacket whose energy is restricted to the discrete part of the spectrum. There are only a few known such systems. Two of them are familiar to any student of quantum mechanics, namely the infinite potential well and the harmonic oscillator, and others include exactly solvable Hamiltonians, in particular P\"oschl-Teller potentials, both in the hyperbolic and trigonometric versions~\cite{Poschl_1933,Loinaz1999}. The question then arises as to whether other potentials with perfect quantum state revivals exist or not.  Here we show that the answer to this question is positive and that there is a vast range of such potentials. To construct them, we employ the intertwining method, and demonstrate quantum state revivals on a number of examples. 

\section{\label{revivals}Quantum State Revivals}

Suppose we have a quantum particle in a static 1D potential $V(x)$ governed by the Schr\"{o}dinger equation $\mathcal{H} \psi(x,t) = E \psi(x,t)$ where the Hamiltonian $\mathcal{H} = -\frac{1}{2}\frac{d^2}{dx^2} + V(x)$ and we work in units in which the mass of the particle and Planck's constant $\hbar$ are equal to unity. To find conditions under which an arbitrary state $\psi(x,t)$ will be periodic in time, we expand the initial wavefunction $\psi(x,0)$ at time $t=0$ as a superposition $\psi(x,0)=\sum_{n}{c_n\psi_n(x,0)}$ of the eigenstates  $\psi_n(x,0)$ of the Hamiltonian. As the wavepacket propagates within the potential, each component acquires a phase factor $\varphi_n(t) =-E_nt$ which depends on its eigenenergy $E_n$, so the state at time $t$ becomes
\begin{equation}
\psi(x,t)=\sum_{n}{c_ne^{-iE_nt}\psi_n(x,0)}\,.
\end{equation}
\noindent For the wavepacket at some moment $t=T_{\mathrm{rev}}$ (the revival time) to be equivalent to that at $t=0$, all the exponential terms $e^{-iE_nt}$ have to be in phase at the time $T_{\mathrm{rev}}$.  This occurs when the energy levels $E_n$ have integral spacing, with a common real factor $a$ and real offset $b$,
\begin{equation}
E_n=aN_n+b\,,
\label{eq:energy}
\end{equation}
and the only requirement on $N_n$ is that it be an integer. The state revival period can then easily be identified as $T_{\mathrm{rev}}=2\pi/a$; at this moment $\psi(x,T_{\mathrm{rev}}) = e^{-\frac{2\pi ib}{a}}\psi(x,0)$, so the wavepacket is revived perfectly up to an unimportant global phase factor. A useful measure of the wavepacket revivals is the autocorrelation function $A(t)=\langle\psi(0)|\psi(t)\rangle$. For a perfect revival, $|A(T_{\mathrm{rev}})|=1$.

It may happen that the Hamiltonian has a spectrum that is partly discrete and partly continuous. If the discrete levels obey the rule (\ref{eq:energy}) and a wavepacket is created only from the eigenstates corresponding to the discrete levels, then the quantum revivals will still occur.

The required energy spacing of (\ref{eq:energy}) is satisfied for the infinite potential well, the quantum harmonic oscillator and its isospectral variants~\cite{Sukumar_1985, Nasuda_2024}, and the P\"{o}schl-Teller potentials, so these systems exhibit perfect wavepacket revivals. In the following we show how to design other potentials with such properties by engineering their energy levels to have suitable patterns through the intertwining technique~\cite{Anderson_1991, Fernandez_2001, Sukumar_1985}. This enables us to obtain an unlimited set of potentials where perfect revivals occur.

\section{\label{constructing}Constructing potentials with desired spectra}

In the following we present an adaptation of the standard intertwining procedure \cite{Anderson_1991, Fernandez_2001, Sukumar_1985, JunkerAndRoy}. The technique begins with a known potential $V_0(x)$ and associated Hamiltonian,
\begin{equation}
 \mathcal{H}_0 = -\frac{1}{2}\frac{d^2}{dx^2}+V_0(x)\,.
\label{eq:Hamiltonian}
\end{equation} 
A new Hamiltonian $\mathcal H_1$ corresponding to a potential $V_1(x)$ shall be constructed such that it has all the energy levels that $\mathcal H_0$ has, plus an additional level that shall lie below all those of $\mathcal H_0$.  The total number of bound states of $\mathcal H_1(x)$ is thus one greater than in $\mathcal H_0$. If $V_0(x)$ is one of the potentials whose discrete levels obey the constraints of (\ref{eq:energy}) and we then add a new energy level below its ground state at the right energy, a new potential retaining the quantum state revival property will result. It is then possible to iterate the intertwining procedure with $V_1(x)$ as the starting point to add another energy level below its ground state, and then continue to create as many new energy levels as desired. 

To find $V_1(x)$, we first factorize the Hamiltonian $\mathcal{H}_0$ as
\begin{align}
\label{eq:H0}
\mathcal{H}_0 = A^{+} A^{-} + \mathscr{E}\,,\qquad
\end{align}
where $A^{\pm}$ are mutually Hermitian-conjugated operators in the form
\begin{equation}
A^{\pm}=\frac{1}{\sqrt{2}}\left[\pm\frac{d}{dx}+U_1(x)\right]
\label{eq:A}
\end{equation}
and $U_1(x)$ is a function to be determined. The factorization (\ref{eq:H0}) is possible if the value of $\mathscr{E}$ does not exceed the ground state energy $E_0^{(0)}$ of the Hamiltonian $\mathcal{H}_0$ because the operator $A^{+} A^{-}=\mathcal{H}_0- \mathscr{E}$ must be positive semidefinite. In the following we will assume that $\mathscr{E}<E_0^{(0)}$, and $\mathscr{E}$ will play the role of the new energy level to be added. Upon substituting Eqs.~(\ref{eq:A}) and ~(\ref{eq:Hamiltonian}) into Eq.~(\ref{eq:H0}) while taking care that the operators are handled correctly, we get an  equation for $U_1(x)$
\begin{equation}
U'_1(x)+U_1^2(x)=2[V_0(x)-\mathscr{E}]\,,
\label{eq:Ricatti}\end{equation}
where prime denotes derivative with respect to $x$. Equation~(\ref{eq:Ricatti}) is known as Riccati equation \cite{Riccati1724,Nieto_1984}.

The new Hamiltonian with the added level $\mathscr{E}$ will then be
\begin{align}
\label{eq:H1}
\mathcal{H}_1 &= A^{-} A^{+} + \mathscr{E}\,.
\end{align}
To show that $\mathcal{H}_1$ has the required properties, we have to show that (i) all the energy levels of $\mathcal{H}_0$ are also energy levels of $\mathcal{H}_1$, and (ii) $\mathcal{H}_1$ has an additional level $\mathscr{E}$ that $\mathcal{H}_0$ does not have. 

To show (i), we denote by $\psi$ a normalized eigenstate of the Hamiltonian  $\mathcal{H}_0$ with energy $E$.  Then Eq.~(\ref{eq:H0}) yields $A^{+} A^{-}\psi=(E-\mathscr{E})\psi$.  Multiplying from the left by $A^{-}$ results in $A^{-} A^{+} (A^{-} \psi)= (E -\mathscr{E})(A^{-} \psi)$, or, equivalently,
 \begin{equation}
 \mathcal{H}_1(A^{-} \psi)= E(A^{-} \psi)\,,
\label{eq:schrodinger1}
\end{equation} 
where Eq.~(\ref{eq:H1}) was used. The square of the norm of the function $A^{-} \psi$ can readily be calculated as $\|A^{-}\psi\|^2 = \langle A^{-}\psi | A^{-}\psi \rangle = \langle \psi | A^{+} A^{-} | \psi \rangle = E - \mathscr{E} > 0$, which can be used to normalize the function $A^{-} \psi$ as $\psi_1\equiv (A^{-} \psi)/\sqrt{E-\mathscr E}$. From Eq.~(\ref{eq:schrodinger1}) we then see that $\mathcal{H}_1\psi_{1}= E\psi_{1}$, i.e., the new Hamiltonian $\mathcal{H}_1$ has a normalized wavefunction $\psi_1$ with energy $E$, so it shares the eigenvalue $E$ with the old Hamiltonian $\mathcal{H}_1$. 

To show (ii), we take advantage of the fact that Eq.~(\ref{eq:Ricatti}) has the solution of the form \cite{Sukumar_1985, Nieto_1984}
\begin{align}
U_1(x) &=\frac{d}{dx}\left[\ln\psi_0(x)\right]=\frac{\psi'_0(x)}{\psi_0(x)}
 \,,
\label{eq:U1}\end{align}
where the function $\psi_0(x)$ satisfies 
\begin{equation}
 \psi_0''(x)+2[\mathscr{E}-V_0(x)]\psi_0(x)=0\,.
\label{eq:schrodinger2}\end{equation}
This can easily be verified by substituting Eq.~(\ref{eq:U1}) into Eq.~(\ref{eq:Ricatti}) and employing Eq.~(\ref{eq:schrodinger2}). At the same time, Eq.~(\ref{eq:schrodinger2}) is nothing else than the stationary Schr\"odinger equation $\mathcal{H}_0\psi_0=\mathscr{E}\psi_0$. The function $\psi_0$ cannot be normalizable, however, since $\mathscr{E}$ does not belong to the spectrum of $\mathcal{H}_0$; as we have assumed, $\mathscr{E}$ lies below the ground state energy of $\mathcal{H}_0$. Still, $\psi_0$ turns out to be very useful because it can be used for constructing the desired normalizable eigenstate of $\mathcal{H}_1$ with eigenvalue $\mathscr{E}$.

To do that, we realize that the function $\psi_0$ must be nodeless, otherwise  Eq.~(\ref{eq:U1}) would lead to a singular function $U_1(x)$. A general solution of Eq.~(\ref{eq:schrodinger2}) contains two arbitrary constants that can be chosen as $\alpha=\psi'_0(0)$ and  $\beta=\psi_0(0)$. Equation~(\ref{eq:U1}) further reveals that multiplying $\psi_0$ by a constant does not change $U_1$, so without loss of generality we can assume that $\beta=1$. 

It is convenient to split $\psi_0(x)$ into even and odd parts, $\psi_0(x)=\psi_{\mathrm{even}}(x)+\alpha\psi_{\mathrm{odd}}(x)$, with the initial conditions $\psi'_{\mathrm{even}}(0)=0, \psi_{\mathrm{even}}(0)=1$ and $\psi_{\mathrm{odd}}(0)=0,\psi'_{\mathrm{odd}}(0)=1$. The fact that  $\mathscr{E}$ lies below the ground state of $\mathcal{H}_0(x)$ ensures that $\psi_{\mathrm{even}}(x)$ be free of zeros. If $\psi_{\mathrm{odd}}(x)$ is included in the solutions of $\psi_0(x)$, care must be taken that $|\alpha|$ is not too large, otherwise the zero present in $\psi_{\mathrm{odd}}(x)$ would show up in $\psi_0(x)$. On the other hand, if $\alpha$ is chosen appropriately, then $\psi_0(x)$ is positive for all $x$, which then yields a non-singular, real function $U_1(x)$ via Eq.~(\ref{eq:U1}). Moreover, due to the fact that  $\mathscr{E}$ lies below the ground state of $\mathcal{H}_0(x)$, the function $\psi_0(x)$ will have an exponentially growing behavior for $x\to\pm\infty$; the only exception would be choosing a particular positive value of $\alpha$ such that $\psi_0(x)$ would converge to zero for $x\to-\infty$ (but still diverge for $x\to\infty$), or choosing a particular negative value of $\alpha$ such that $\psi_0(x)$ would converge to zero for $x\to\infty$ (but still diverge for $x\to-\infty$). Excluding such values from the possible choices of $\alpha$, we arrive at a positive real function $\psi_0(x)$ that has an exponentially growing behavior for $x\to\pm\infty$. Now consider a function $\psi_{01}(x)=1/\psi_0(x)$; it is normalizable due to the exponentially growing behavior of $\psi_0(x)$. A simple calculation further reveals that 
\begin{align}
 \sqrt2\,A^+\psi_{01}=\psi'_{01}+U_1\psi_{01} =-\frac{\psi'_{0}}{\psi^2_{0}}+\frac{\psi'_{0}}{\psi_{0}}\frac1{\psi_{0}}=0\,.
\label{}\end{align}
From this it follows that $A^-A^+\psi_{01}=(\mathcal H_1-\mathscr{E})\psi_{01}=0$; in other words, the Hamiltonian $\mathcal H_1$ has a normalizable wavefunction with the eigenvalue $\mathscr{E}$, which completes the part (ii).

The last step to be done is expressing the potential $V_1(x)$ in the new Hamiltonian $\mathcal H_1$. This can easily be performed by combining Eqs.~(\ref{eq:H1}) and~(\ref{eq:A}), which yields 
\begin{align}
\nonumber
\mathcal{H}_1 &=\frac12 \left[-\frac{d}{dx}+U_1(x)\right]\left[\frac{d}{dx}+U_1(x)\right]+\mathscr{E}\\
 &= -\frac{1}{2}\frac{d^2}{dx^2}+V_1(x)\,,
\end{align}
where the new potential is
\begin{equation}
  V_1(x)=\frac12\,[U_1^2(x)-U'_1(x)]+\mathscr{E}=U_1^2(x)-V_0(x)+2\mathscr{E}
\label{eq:V1}\end{equation}
and Eq.~(\ref{eq:Ricatti}) has been used. Alternatively, $V_1(x)$ could be expressed in terms of the function $\psi_0$, which leads to
\begin{equation}
  V_1(x)=V_0(x)-\frac{d^2}{dx^2}\,\ln\psi_0\,.
\label{eq:V1-alt}\end{equation} 
This is the form derived in Ref.~\cite{Sukumar_1985}. It reveals that due to the properties of the function $\psi_0(x)$ discussed above, $V_1(x)$ is a real and well behaved function. 

For a practical calculation of the new potential $V_1(x)$, in particular a numerical calculation, it is more convenient to use Eq.~(\ref{eq:V1}) rather than  Eq.~(\ref{eq:V1-alt}) because one then avoids the astronomically large values that the unnormalizable $\psi_0(x)$ contains. The $U_1(x)$ is found by solving Eq.~(\ref{eq:Ricatti}), for which an initial condition is also needed; it follows from the initial conditions for $\psi_0(x)$ and  has the form $U_1(0)=\alpha$. Provided that the original potential is symmetric, i.e., $V_0(-x)=V_0(x)$, the choice $\alpha=0$ yields an antisymmetric (odd) function $U_1(x)$ and consequently a symmetric potential $V_1(x)$; in the following we will restrict ourselves to this case. 

\begin{algorithm}
\caption{Procedure for Creating Potential with $n$ New Energy Levels}
\KwData{%
    \begin{itemize}
        \item Starting symmetric potential $V_0(x)$
        \item A decreasing sequence of desired new levels $\mathscr{E}_1,\dots,\mathscr{E}_n$
    \end{itemize}
}
\KwResult{New potential $V_n(x)$}
        Solve the equation for $U_1(x)$:\\
        $U'_1(x)+U_1^2(x)=2[V_0(x)-\mathscr{E}_1]$\\ 
        with $U_1(0) = 0$.\\
    \For{$i \gets 2$ \KwTo $n$}{
        Solve the equation for $U_i(x)$:\\
\Indp
$U'_{i}+U^2_{i}=-U'_{i-1}+U^2_{i-1}+2(\mathscr{E}_{i-1}-\mathscr{E}_{i})$ \\
with $U_i(0) = 0$.\\
\Indm
\Indp
\Indm 
       }
    \KwRet $V_n=(U_n^2-U_n')/2+\mathscr{E}_{n}$\;
Output the final potential containing the additional $n$ energy levels.\\
\end{algorithm}

Equation~(\ref{eq:V1}) provides the desired new potential $V_1(x)$ with the added level $\mathscr{E}$ and it is suitable for numerical calculations. The entire procedure can be iterated to add arbitrarily many new energy levels, provided that each new level be below all of the previously added levels. This way, if $n$ new levels $\mathscr{E}_1,\dots,\mathscr{E}_n$ are being added consecutively, one obtains a sequence of functions $U_1,\dots,U_n$ and potentials $V_1,\dots,V_n$. However, for a practical calculation when only the last potential $V_n$ is of interest, it is more convenient to work directly with the functions $U_i$ and avoid the functions $V_i$ in the intermediate steps. For this purpose, one can derive a recurrence equation for the functions $U_i$ in the form
\begin{equation}
  U'_{i}(x)+U^2_{i}(x)=-U'_{i-1}(x)+U^2_{i-1}(x)+2(\mathscr{E}_{i-1}-\mathscr{E}_{i})
\label{eq:UviaU}\end{equation}
for $ i=2,3,\dots,n$. For calculating $U_1$, Eq.~(\ref{eq:Ricatti}) is used with $\mathscr{E}$ replaced by $\mathscr{E}_1$, and the final potential is obtained as 
\begin{equation}
 V_n(x)=\frac12[U_n^2(x)-U_n'(x)]+\mathscr{E}_n
\label{eq:Vn}\end{equation}
in analogy with Eq.~(\ref{eq:V1}). Algorithm 1 summarizes the whole procedure. We have tested the algorithm to construct potentials with various spectral properties; a number of examples are shown in the next section. While it is known that this method can be used to add energy levels to existing potentials (e.g., a single level insertion is illustrated in Ref. ~\cite{Sukumar_1985}), we have found that hundreds of levels can be added easily when working with $U_i$.

Aside from this caveat, however, the intertwining method itself is well known, with a formalism similar to that of supersymmetric quantum mechanics~\cite{Cooper1995} which emerged from Witten’s pioneering work~\cite{Witten1981} as a simplified framework to study symmetry breaking in quantum field theory, later revealing deep connections with solvable quantum potentials.  
In this formalism, \( A^- \) and \( A^+ \) are known as \emph{supercharge operators}  and \( U_1(x) \) is the \emph{superpotential}.  The partner Hamiltonians \( H_0 \) and \( H_1 \) are said to be \emph{intertwined} by these operators. Historically, these constructions trace back to Darboux’s transformation~\cite{Darboux1882}, which described how one linear differential equation could generate another with shifted spectral properties; this was later systematized in quantum mechanics~\cite{Infeld1941,Gendenshtein1983}. Mielnik~\cite{Mielnik1984} generalized the factorization to yield a continuous set of isospectral deformations, and showed this in depth with the example of the quantum harmonic oscillator.  Further developments by Andrianov \emph{et al.}~\cite{Andrianov1984} and Bagrov and Samsonov~\cite{Bagrov1995} established the correspondence between Darboux transformations, intertwining operators, and supersymmetric partner Hamiltonians.  These works showed that successive Darboux transformations could generate new potentials with selectively added or removed bound states, which we employ precisely for such spectral engineering. This demonstrates the applicability of supersymmetry far beyond its original intentions. Indeed, it can be said that the very existence of potentials with identical spectra is due to hidden supersymmetry~\cite{Andrianov1984}.

The intertwining method works, by its nature, in the ``top to bottom'' direction: we start from a given potential and add new energy levels below its existing levels. However, it might be more natural to work ``bottom to top'', starting with the ground state and then adding sequentially higher energy levels. Fortunately, the method is suitable for this purpose too: We first list as many levels as desired that follow a required ``bottom to top'' pattern in our new potential. We then choose any suitable initial potential, typically the harmonic oscillator or a constant potential; we then invert our list of desired levels and add them in reverse order.

\subsection{Influence of the choice of $V_0(x)$ on $V_n(x)$}

It is natural to ask what is the influence of the choice of the original potential $V_0(x)$ on the resulting final potential $V_n(x)$ for given energy levels $\mathscr{E}_1,\dots,\mathscr{E}_n$ to be added. While it turns out that the choice of $V_0$ influences $V_n$ in complex ways, interestingly, the value of $V_n$ at the origin can be calculated analytically. By combining Eq.~(\ref{eq:Ricatti}), Eqs.~(\ref{eq:UviaU}) for $i=2,\dots,n$ and Eq.~(\ref{eq:Vn}), and employing the fact that $U_i(0)=0$ for all $i=1,\dots,n$, we arrive at
\begin{equation}
V_n(0)=(-1)^n\,V_0(0)+2\sum_{i=1}^n(-1)^{n-i}\,\mathscr{E}_{i}\,.
\label{eq:Vn0}\end{equation}
This equation shows that if we keep the added levels $\mathscr{E}_1,\dots,\mathscr{E}_n$ fixed and increase the value $V_0(0)$
[e.g., by shifting the whole potential $V_0(x)$], the value $V_n(0)$ increases by the same amount if $n$ is even while it decreases by the same amount if $n$ is odd. Consequently, choosing the original potential $V_0$ inappropriately can lead to strong oscillations of the potential $V_n$. 

It is in principle possible to evaluate also the derivatives $V_n^{(k)}(0)$ analytically; for odd $k$, they are identically zero while for even $k$ they lead to increasingly complex expressions involving the added energy levels and the derivatives of the original potential $V_0$ at the origin.

\section{\label{constructing_revivals}Constructing potentials with perfect revivals}

In the following we introduce several examples of energy level structure that yield perfect quantum state revivals, and construct the corresponding potentials. We also demonstrate the revivals using the autocorrelation function and quantum carpets.

\subsection{Biperiodic oscillator}

\begin{figure}[tbh]
\includegraphics[width=0.9\columnwidth]{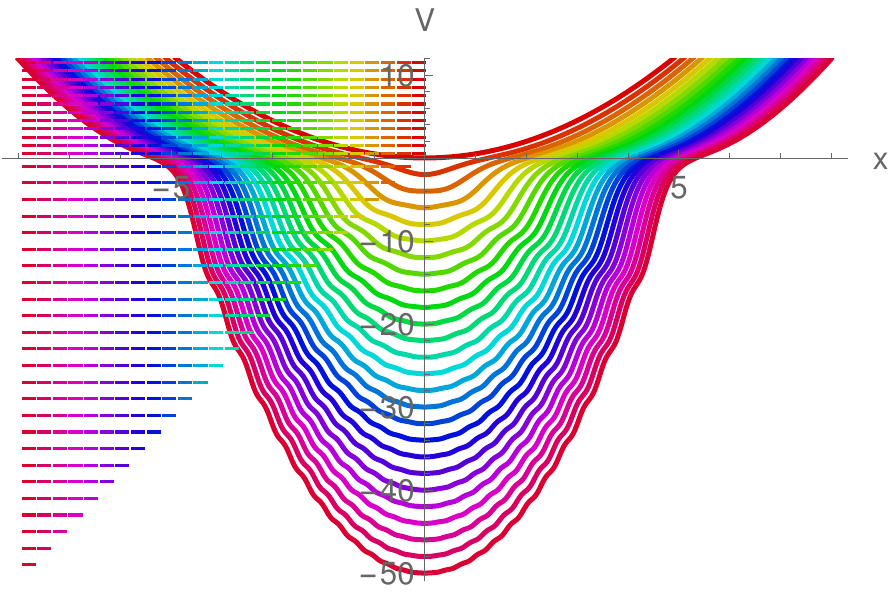}
\caption{\label{fig:biperiodic1} Biperiodic potentials. Starting from the harmonic potential $x^2/2$ (the uppermost curve, red), $N=25$ additional energy levels with spacing of 2 at values $-1,-3,\dots,-49$ have been added successively using the scheme in Algorithm 1, creating 25 new potentials. The numerically calculated spectra of each potential are shown towards the left in matching colors; the leftmost column shows the spectrum of the lowermost potential. The fact that the levels of various potentials match is apparent.}
\end{figure}

Let's start from the harmonic oscillator potential with energy levels $E_n=n+\frac{1}{2}$ for $n\geq 0$ and sequentially add an additional $N$ levels $E = -1, -3, -5, \ldots, -(2N-1)$. Employing the semiclassical relation between separation of quantum energy levels and classical period of oscillations~\cite{ROBINETT_2004}, we find that this would correspond to a classical biperiodic oscillator that has frequency $\omega_1=1$ for $E>0$ and  $\omega_2=2$ for $E<0$; such a potential was described for instance in Ref.~\cite{Tyc_2015}. The resulting quantum potentials and energy levels are shown in Fig.~\ref{fig:biperiodic1} for $N$ ranging from 1 to 25; the levels were numerically calculated from the constructed potentials and they match the desired levels with a high precision. The quantum potentials closely resemble their classical counterpart (see Fig.~1(a) in Ref.~\cite{Tyc_2015}). We can conveniently represent the time evolution and revival of the wavepacket by the quantum carpet~\cite{ROBINETT_2004}, which is analogous to the Talbot carpet in optics~\cite{Berry2001}. Such a carpet represents the time evolution of a wavepacket $\psi(x,t)$ as a 2D plot of the probability density $|\psi(x,t)|^2$, where the time variable is on the vertical axis and coordinate on the horizontal axis. Perfect wavepacket revivals manifest themselves by the periodicity of the carpet in the time direction.

\begin{figure}[htbp]
\includegraphics[width=0.9\columnwidth]{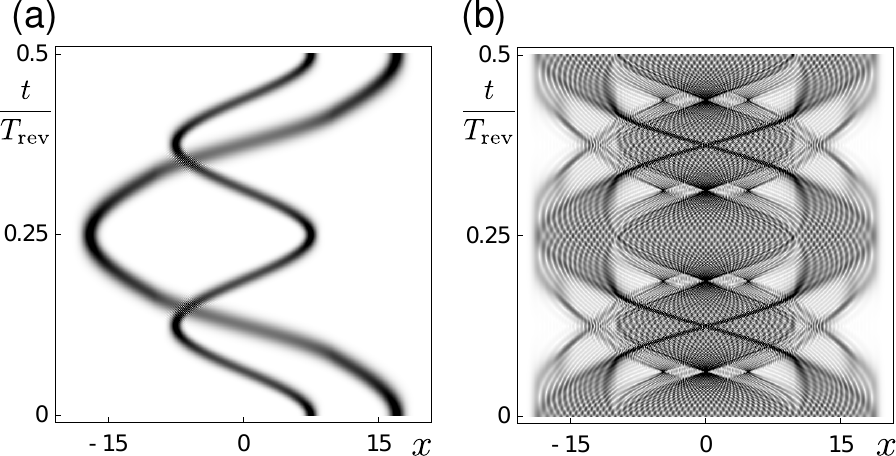}
\caption{\label{fig:talbot-biperiodic1} Quantum carpets for the biperiodic potential where 100 levels with spacing of 2 were added to the original harmonic potential $x^2/2$. The absolute value of the wavefunction is indicated by the brightness (larger value corresponds to a darker color), the vertical axis represents time proportional to $T_{\mathrm{rev}}=4\pi$. Two initial wavepackets were used: (a) superposition of two Gaussian states and (b) a shifted cosine oscillating between 0 and 1. It is apparent, especially in (a), that the period of the lower-amplitude Gaussian component is equal to half of the period of the higher-amplitude component.}
\end{figure}

\begin{figure}[htbp]
\includegraphics[width=0.9\columnwidth]{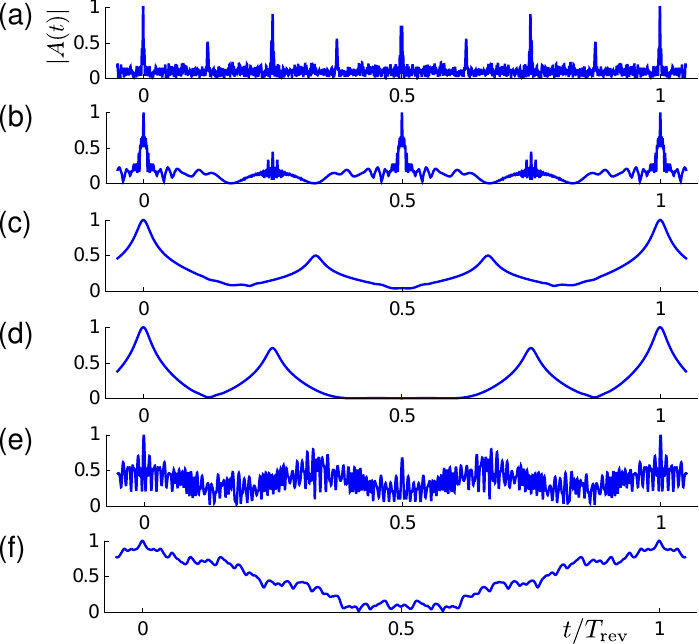}
\caption{\label{fig:correlation} The plots show the magnitude of the autocorrelation function, $|A(t)|$, for the following quantum carpets: (a) the carpet from Fig.~\ref{fig:talbot-biperiodic1}(b), (b) the carpet from Fig.~\ref{fig:talbot-biperiodic2}(b), (c-d) the carpets from Fig.~\ref{fig:alternating}(e-f), respectively, (e,f) the carpets from Fig.~\ref{fig:primes-50}(b,d), respectively.}
\end{figure}

Fig.~\ref{fig:talbot-biperiodic1} shows examples of quantum carpets that represent the time evolution of wavepackets for the biperiodic potential with $N=100$ added levels, where the bi-periodicity is apparent. 
Fig.~\ref{fig:correlation}(a) shows the magnitude of the autocorrelation function $|A(t)|$ and confirms perfect wavepacket revivals.

\subsection{Reverse biperiodic oscillator}

\begin{figure}[htbp]
\includegraphics[width=0.9\columnwidth]{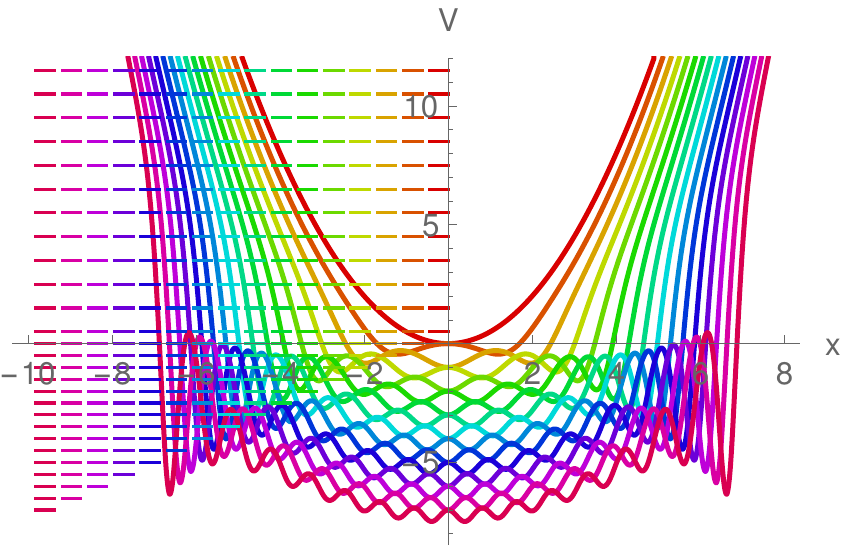}
\caption{\label{fig:biperiodic2} Same as Fig.~\ref{fig:biperiodic1}, but for reverse biperiodic potentials.  $N=15$ levels were added successively with spacing of $1/2$ at values $0,-1/2,-1,\dots,-7$. Fewer added levels than in Fig.~\ref{fig:biperiodic1} were used for the sake of clarity of the plot.}
\end{figure}

Consider now a situation similar to the previous one: we again start from the harmonic oscillator potential with energy levels $E_n=n+\frac{1}{2}$ for $n\geq 0$ and sequentially add an additional $N$ levels $E =0, -\frac12, -1, -\frac32, \ldots, -(N-1)/2$, this time with a~smaller separation than the original levels. The resulting quantum potentials and energy levels are shown in Fig.~\ref{fig:biperiodic2}.

\begin{figure}[htbp]
\includegraphics[width=0.9\columnwidth]{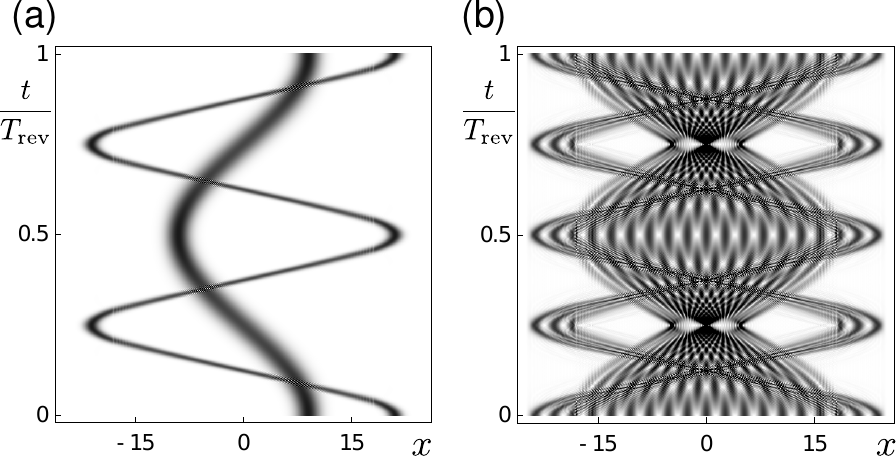}
\caption{\label{fig:talbot-biperiodic2} Quantum carpets for the reverse biperiodic potential where 100 levels with spacing of 1/2 were added to the original harmonic potential $x^2/2$. The initial conditions are similar as in Fig.~\ref{fig:talbot-biperiodic1}. The biperiodicity is apparent again, this time the period of the lower-amplitude component equals twice the period of the higher-amplitude component. The revival time is $T_{\mathrm{rev}}=4\pi$.}
\end{figure}

Classically, this spectrum would correspond to a biperiodic oscillator with frequency $\omega_1=1$ for $E>0$ and  $\omega_2=1/2$ for $E<0$. It turns out, however, that such a classical potential does not exist. Indeed, when solving the inverse problem of finding a classical potential $V(x)$ from a given dependence $T(E)$ of period on energy~\cite{landau1991mechanics}, one first obtains the function $x(V)$ that has to be inverted to get $V(x)$; since the resulting function is decreasing here, this is impossible. Such a non-existence of the classical counterpart is related to the oscillations of the quantum potentials apparent in Fig.~\ref{fig:biperiodic2}; the particle tunnels between local potential minima, which is not possible classically. Quantum carpets for the reverse biperiodic potential with $N=100$ added levels are shown in Fig.~\ref{fig:talbot-biperiodic2}, the bi-periodicity is apparent again. The autocorrelation function is shown in Fig.~\ref{fig:correlation}(b).

\subsection{Potentials with alternating gaps between levels}

As the next example, let us construct a potential with the gap between the subsequent levels alternating between two values. Starting from the harmonic oscillator, we construct $N=100$ additional levels following this pattern. 

\begin{figure}[htbp]
\includegraphics[width=0.9\columnwidth]{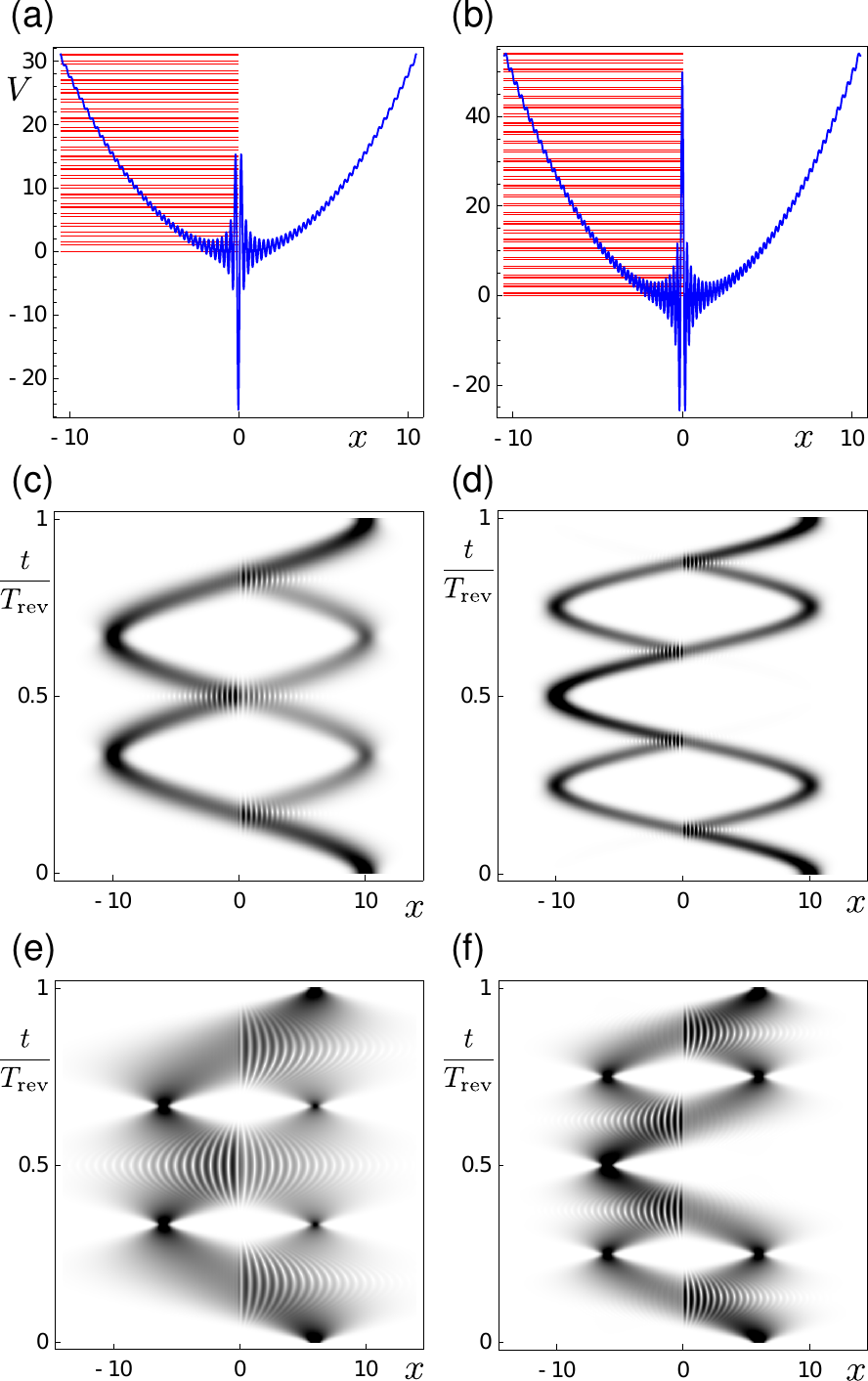}
\caption{\label{fig:alternating}  Potentials with alternating gaps between the energy levels. The gaps alternate (a) between the values 1 and 1/2 and (b) between the values 1/2 and 3/2. The potentials were shifted by suitable constants to produce zero ground state energy. The spectrum is shown by the red lines. (c-d) show a quantum carpet for the potentials in (a-b), respectively, for an initial Gaussian wavepacket. (e-f) the same as (c-d), but for a narrower initial wavepacket. The potentials seem to act like diffraction gratings for the incoming wavepackets, splitting them into two and then rejoining them into one again. In both cases, the revival time is $T_{\mathrm{rev}}=4\pi$.}
\end{figure}

The resulting potentials are illustrated in Fig.~\ref{fig:alternating}(a-b) for two particular cases of gap patterns between the levels. Fig.~\ref{fig:alternating}(c-f) then shows the corresponding quantum carpets exhibiting an interesting quantum interference effects. The autocorrelation functions are shown in Fig.~\ref{fig:correlation}(c,d).

\subsection{Potentials with prime and Fibonacci spectra}

\begin{figure}[htbp]
\includegraphics[width=0.9\columnwidth]{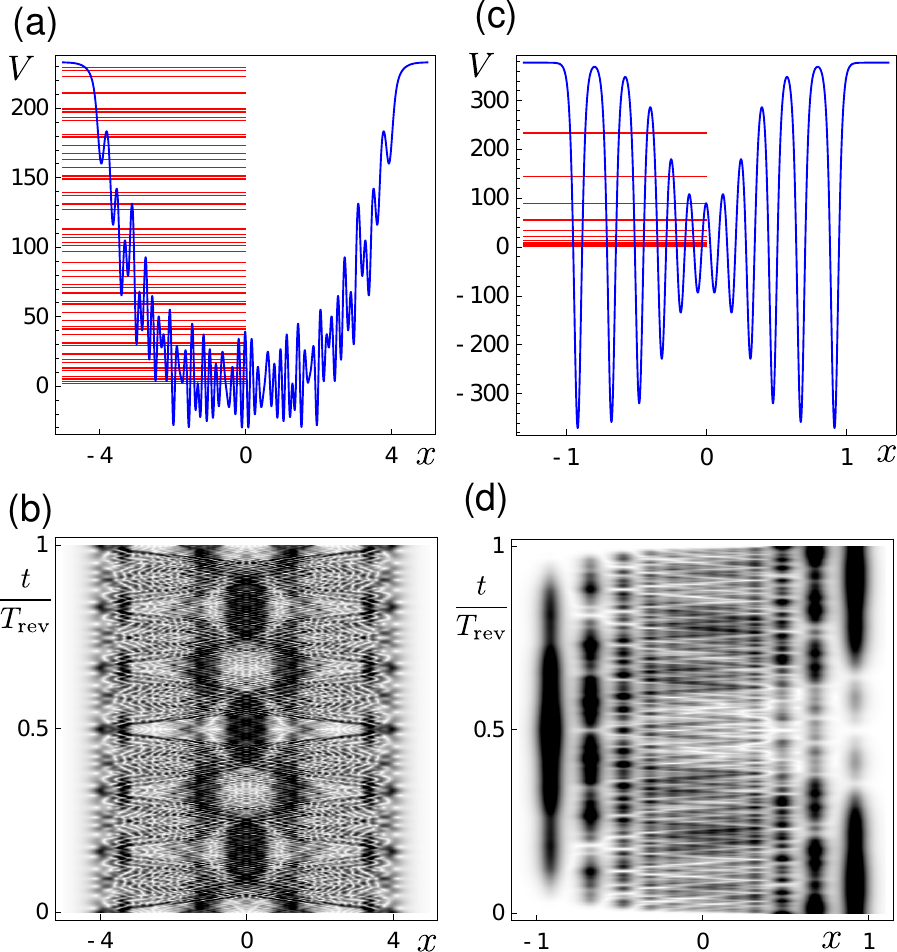}
\caption{\label{fig:primes-50} (a) Potential with spectrum given by the first 50 prime numbers ($2,3,5,\dots,229$) and (b) its quantum carpet. (c) Potential with spectrum given by the 12 Fibonacci numbers $1,2,3,5,8,\dots,233$ and (d) its quantum carpet. In both cases, the revival time is $T_{\mathrm{rev}}=2\pi$.}
\end{figure}

To demonstrate the power of the method, we present two more examples, namely potentials whose spectra are given by prime and Fibonacci numbers. Denoting the $n^{\mathrm{th}}$ prime number by $p_n$, we start from the constant potential at the level $p_{N+1}$ for some $N\in\mathbb N$ and add sequentially $N$ levels $p_N,p_{N-1},\dots,p_1$. Fig.~\ref{fig:primes-50}~(a--b) shows the resulting potential for $N=50$ and an example of the quantum carpet; the autocorrelation functions is shown in Fig.~\ref{fig:correlation}(e). Next we repeat the same procedure for Fibonacci numbers $F_n$ instead of primes; however, the first Fibonacci number 1 has to be omitted to avoid repeating level values, so the designed level values are $F_2,F_3,\dots,F_N$. The resulting potential has a regular structure; it is depicted along with its quantum carpet in Fig.~\ref{fig:primes-50}~(c--d) for $N=13$; the autocorrelation function is shown in Fig.~\ref{fig:correlation}(f).

\section{Error analysis}
\label{erroranalysis}

\begin{figure}[htb]
\includegraphics[width=\columnwidth]{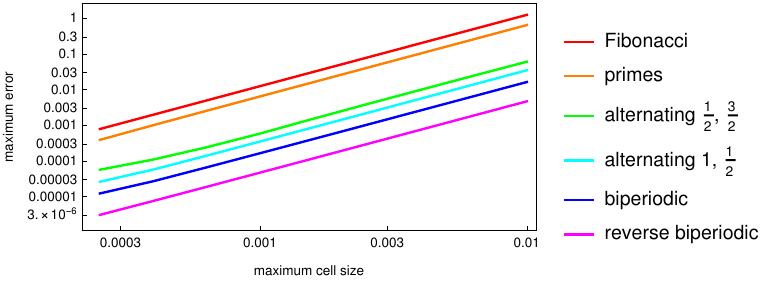}
\caption{\label{fig:errors} The dependence of maximum errors of actual energy levels on the maximum cell size used in the finite element method for the particular potentials discussed in Sec.~\ref{constructing_revivals}.}
\end{figure}

Naturally, it is of interest to know the level of precision with which the actual energy levels of the potentials obtained by our method match the desired levels used to design them. We have performed such analysis for the six potentials discussed in the previous section. To calculate the energy spectrum, we have implemented the finite element method where the maximum cell size (the variable ``MaxCellMeasure'' in the software Mathematica) can be chosen. For various values of this variable we have then calculated the energy spectrum, compared it with the desired energy levels, and evaluated the maximum error. The results are shown in Fig.~\ref{fig:errors}. The best results were obtained for the reverse biperiodic and biperiodic potentials while the worse ones were obtained for the prime and Fibonacci spectra. As a general rule, in the log-log scale all the curves have a slope of approximately 2, meaning that decreasing the maximum cell size by an order of magnitude results in the maximum error reduction by two orders of magnitude. This behavior suggests that the errors are not caused so much by the potentials themselves, but rather by the procedure of calculating the levels. This way, our method enables to design  potentials that have the desired energy levels with a high degree of precision.

\section{Multidimensional case}

Our method can be readily extended to multiple dimensions. Suppose we have designed potentials $\{V_k(x);k=1,\dots,n\}$, all with the property (\ref{eq:energy}) with common values of $a$ and $b$. Consider now a combined potential $V(x_1,\dots,x_n)=\sum_{k=1}^n V_k(x_k)$, where $x_k$ are Cartesian coordinates of the $n$-dimensional Euclidean space. The energy levels of a particle moving in the potential $V$ will be in the form of sums of energy levels of the individual potentials $V_k$, as can easily be shown via separation of variables in the $n$-dimensional Schr\"odinger equation. Therefore the energy levels in the potential $V$ will also have the property~(\ref{eq:energy}) and will lead to perfect revivals of $n$-dimensional wavepackets. 

\vfill
\section{Conclusions}

We have presented a general method for designing potentials that exhibit perfect quantum state revivals. The examples given demonstrate the power and flexibility of the method; there exist  endless possibilities for other energy level patterns that obey  Rule (\ref{eq:energy}); constructing their corresponding quantum potentials leads to a multitude of potentials that exhibit perfect wavepacket revivals.  The intertwining method described above and applied in Algorithm 1 can also be used to quickly and easily design potentials that have any desired energy level positions and is thus a powerful inverse design technique that may possibly have applications beyond what we have considered here.  Since separation of variables allows the construction of $n$-dimensional potentials from individual 1-dimensional potentials, the spectral engineering of any desired potential is conceivable with the algorithm presented.  Finally, it should be pointed out that the operators $A^+$ and $A^-$ are not unique~\cite{Mielnik1984}; higher-order operators would allow  isospectral deformations similar to what we have shown, but with additional freedom~\cite{Andrianov1993,Fernandez1998,Samsonov1999}. Along with consideration of higher-order intertwining operators, a possible extension of our work could include designing potentials for relativistic quantum revivals~\cite{prl104-120403-strange}.


\providecommand{\noopsort}[1]{}\providecommand{\singleletter}[1]{#1}%
%



\end{document}